# YARDSTICK COMPETITION AND SPATIAL SPILLOVERS OF ECONOMIC INFRASTRUCTURE: A SPATIAL PANEL ANALYSIS OF CAMEROONIAN REGIONS

Emérite-Marc MOUAHA*[1], stemerite@gmail.com

Fidel Djongreba NDIKWA[1,2,] fidel.djong@yahoo.fr

Barnabe ABBA YADOU[3] abbayadou@gmail.com

[1] Department of operational research, Higher Institute of Transport Logistics and commerce, University of Ebolowa, Ebolowa, Cameroon

[2] Department of Mathematics, University of Caen-Normandie, Caen, France

[3] Department of Public Economics, Faculty of Economics and Management, University of Maroua, Maroua, Cameroon

## Abstract

This article tests the theory of "top-down" yardstick competition induced by the central government by examining spatial interactions among Cameroon's ten regions in the allocation of economic infrastructure spending, primarily on roads. Using the SPREGDPD estimator (Arellano–Bond for spatial panels), we estimate a spatial autoregressive (SAR) model on quarterly panel data covering 2015–2018 (160 observations) drawn from Cameroonian Treasury accounts. The results show: (i) a positive and statistically significant spatial interaction between neighbouring regions ($\rho = 0.0348$ to $0.0436$, $p < 0.05$), consistent with the presence of yardstick competition; (ii) a positive effect of own-source revenues on economic infrastructure (marginal effect of 2.8%), confirming the importance of local fiscal autonomy; and (iii) a negative effect of the General Decentralization Grant (−13.5%), suggesting crowding-out or inefficient allocation of unconditional transfers. These findings support policy measures aimed at strengthening local budgetary autonomy, linking part of intergovernmental transfers to performance indicators, and designing targeted instruments for the poorest regions.

## 1. Introduction

A fundamental question that has received limited attention as Cameroon continues its fiscal decentralization process initiated in the early 2000s and reinforced by the General Code of Decentralized Territorial Communities (2019) is whether regional governments engage strategically in the allocation of economic infrastructure expenditures.

According to first-generation fiscal federalism theory (Tiebout, 1956; Oates, 1972), decentralization improves allocative efficiency because local governments possess better information about citizens' preferences. A central mechanism underlying this efficiency is interjurisdictional competition: mobile citizens "vote with their feet" by relocating to jurisdictions that offer the most attractive combination of taxes and public services, thereby disciplining local officials (Tiebout, 1956). Salmon (1987) and Besley & Case (1995) formalised a complementary mechanism known as yardstick competition, whereby voters evaluate their local government's performance relative to neighbouring jurisdictions and reward or punish incumbents at the ballot box, generating horizontal strategic interactions.

However, second-generation fiscal federalism (Weingast, 1995; Rodden, 2004; Besley & Coate, 2003) emphasises that these mechanisms require conditions rarely met in developing countries: absence of clientelist capture, adequate citizen information, competitive local elections, and residential mobility. When electoral accountability is weak, the central government can exogenously induce yardstick competition by allocating discretionary transfers on the basis of relative local performance (Shleifer, 1985). This "top-down" competition mechanism offers a viable substitute for imperfect market or electoral discipline.

Recent empirical research supports the relevance of this approach in sub-Saharan Africa. Studies on Ghana show that transfer allocation is shaped by political alignment and relative district performance, creating implicit competition among local governments. Comparative reviews covering several African countries call for greater attention to spatial spillover effects in decentralization evaluations, noting that standard assessments systematically ignore spillovers and thereby bias treatment-effect estimates. Evidence from Uganda and other countries further indicates that local grants are partly conditioned on relative performance indicators, generating spatially correlated competitive pressures.

Models of new economic geography (Krugman, 1991; Martin & Rogers, 1995) additionally highlight the role of transport infrastructure in shaping the spatial distribution of economic activity and in generating interjurisdictional interdependencies. Infrastructure investments in one region affect location decisions and growth prospects in neighbouring regions, reinforcing the need for a spatial analytical framework.

This paper is, to the best of our knowledge, the first to examine spatial interactions among Cameroonian regional governments in the allocation of economic infrastructure expenditure. It contributes to the literature in three ways. First, it provides the first spatial econometric evidence on strategic interactions among Cameroon's decentralised territorial communities. Second, it extends work on "top-down" yardstick competition to the Central African context, where local electoral accountability remains limited. Third, it quantifies the differentiated effects of alternative local financing instruments, yielding concrete implications for the reform of intergovernmental transfers and equalisation policy in Cameroon.

We address three research questions using a balanced panel of Cameroon's ten regions over the period 2015Q1–2018Q4 (160 quarterly observations):

(1) Do regional governments in Cameroon engage in strategic interactions when allocating economic infrastructure expenditure?

(2) Can the observed interactions be interpreted as yardstick competition induced "from above" by the central government?

(3) How do different types of intergovernmental transfers and own-source revenues shape regional economic infrastructure spending?

The remainder of the paper is organised as follows. Section 2 presents the conceptual framework. Section 3 reviews the empirical literature. Section 4 identifies the research gaps and positions our contribution. Section 5 describes the institutional context and data. Section 6 outlines the methodological strategy. Section 7 reports the empirical results and discusses marginal effects. Section 8 derives policy implications. Section 9 concludes.

## 2. Conceptual Framework

The theoretical framework rests on three complementary pillars: (i) fiscal federalism and interjurisdictional competition; (ii) yardstick competition and its adaptation to developing-country contexts; and (iii) new economic geography and the modelling of spatial spillovers.

### 2.1 Fiscal Federalism and Interjurisdictional Competition

The theory of fiscal federalism, initiated by Musgrave (1959) and formalised by Oates (1972), provides the analytical foundation for the economics of decentralization. It distinguishes three main functions of government allocation, redistribution and stabilization and argues that the allocation function is performed more efficiently at the local level, provided that interjurisdictional externalities are internalised. Oates's decentralization theorem states that each public service should be provided by the jurisdiction that most closely matches the geographic extent of its benefits and costs.

Interjurisdictional competition is a key mechanism supporting allocative efficiency. Tiebout (1956) famously argued that mobile and informed citizens “vote with their feet” by choosing the jurisdiction whose tax–service package best matches their preferences. This mobility disciplines local governments and pushes them toward efficient provision. Second-generation fiscal federalism (Weingast, 1995; Besley & Coate, 2003; Rodden, 2004) relaxes the strong assumptions of the first-generation model by incorporating information asymmetries, agency problems and strategic behaviour. In this perspective, local officials subject to weak electoral constraints may engage in rent-seeking or clientelist capture, undermining the expected benefits of decentralization (Bardhan & Mookherjee, 2006).

## 2.2 Yardstick Competition in Developing Countries

Salmon (1987) and Besley & Case (1995) proposed yardstick competition as a complementary discipline mechanism that can operate even in the absence of perfect residential mobility. Voters who are imperfectly informed about the cost and efficiency of local public services compare the performance of their own government with that of neighbouring jurisdictions. Relative under-performance leads to electoral punishment; relative over-performance leads to re-election. This comparative evaluation generates horizontal strategic interactions: each local government takes into account the policy choices of its neighbours because those choices affect the benchmark against which it is judged.

In many developing countries, however, the electoral channel of yardstick competition is weak because of limited residential mobility, low information and imperfect local democracy (Bardhan, 2002; Prud’homme, 1995). Shleifer (1985) suggested an alternative “top-down” mechanism in which the principal (the central government) creates competition among agents (local governments) by linking rewards typically intergovernmental transfers to relative performance. This mechanism requires only that the centre be able to observe and compare local outcomes; it does not require competitive local elections or high citizen mobility. Recent applications of this logic to African settings show that central governments do indeed use discretionary transfers to reward relative performance or political alignment, thereby generating implicit competition among local jurisdictions.

## 2.3 Spatial Spillovers and New Economic Geography

New economic geography (Krugman, 1991) provides a complementary framework for understanding how economic infrastructure, especially roads, generates spatial spillover effects. Reductions in transport costs influence firm location decisions and the formation of agglomerations. Infrastructure investment in one region can attract economic activity from neighbouring regions or, conversely, generate positive demand spillovers. Martin & Rogers (1995) distinguish between intra-regional and inter-regional infrastructure and show that the spatial distribution of public investment can either reinforce or reduce existing regional disparities. These interdependencies justify a spatial econometric approach when analysing regional infrastructure spending.

## 3. Review of the Literature: Fiscal Decentralization, Yardstick Competition and Spatial Interactions

### 3.1 Evidence from Developed Countries

A substantial body of empirical research has documented yardstick competition in developed countries. In their seminal study, Case, Hines and Rosen (1993) show that U.S. state governors adjust public expenditures in response to the spending decisions of neighbouring states, with spatial interaction coefficients typically ranging between 0.2 and 0.4. Similar evidence has been found for Belgian municipalities (Ashworth & Heyndels, 1997), Italian local governments (Bordignon, Cerniglia & Revelli, 2003), English local authorities (Revelli, 2006) and Swiss cantons (Schaltegger & Kittel, 2002). Foucault, Madies and Paty (2008), using French municipal data, provide an important nuance: spatial interactions are significantly stronger for visible capital expenditures (including infrastructure) than for current expenditures. This finding is particularly relevant for the present study, which focuses on economic infrastructure spending.

Overall, the literature on high-income countries consistently finds positive and significant spatial interaction coefficients, generally lying between 0,10 and 0,50 depending on the country, the tier of government and the category of expenditure (Revelli, 2006). These results are commonly interpreted as evidence of yardstick competition driven by comparative electoral evaluation.

### 3.2 Evidence from Developing Countries and Sub-Saharan Africa

In developing countries, the classical electoral channel of yardstick competition is often impaired by limited residential mobility, imperfect information and weak local democracy (Bardhan, 2002; Prud'homme, 1995). Nevertheless, a growing literature documents the existence of spatial interactions among local governments through alternative mechanisms.

Using data from Indonesian districts, Arze, Martinez-Vazquez and Puwanti (2008) find significant spatial dependence in public spending, with an estimated spatial coefficient of approximately 0.15. Chavis (2009) shows that competition for grants among Indonesian villages reduces the unit cost of road construction, providing indirect but compelling evidence of strategic interaction. In the African context, Caldeira, Foucault and Rota-Graziosi (2015) and subsequent work on Benin document spatial interactions in local public spending. More recent studies on Ghana show that intergovernmental transfers are influenced by political alignment and relative performance considerations. Evidence from Morocco (Elkhdari et al., 2021) confirms that municipal expenditure choices are influenced both by geographical neighbours and by municipalities sharing the same political affiliation.

Importantly, several studies suggest that in settings where local electoral accountability is weak, the central government can itself generate “top-down” yardstick competition by conditioning discretionary transfers on relative local performance or political criteria (Shleifer, 1985). This mechanism does not require competitive local elections or high citizen mobility; it only requires that the centre be able to observe and compare local outcomes. The Cameroonian case, characterised by a still limited degree of local fiscal autonomy and a heavy reliance on central transfers, provides a relevant setting in which to test the presence of such top-down interactions.

## 3.3 Spatial Econometrics and the Identification of Strategic Interactions

Spatial econometrics supplies the formal apparatus required to identify strategic interactions. Anselin (1988) distinguishes the spatial autoregressive (SAR) model, the spatial error model (SEM) and the spatial Durbin model (SDM). These frameworks have been extended to panel data by Elhorst (2003, 2005), Kelejian and Prucha (1998, 2004), Kapoor, Kelejian and Prucha (2007) and Lee and Yu (2009, 2010). The SAR specification is particularly well suited to capturing horizontal interjurisdictional interactions because it explicitly includes a spatial lag of the dependent variable. Specification tests Moran’s I, LM-Lag, LM-Error and their robust versions guide the choice among alternative spatial models.

Recent comparative work confirms the usefulness of the SAR model for detecting strategic interactions in public spending. Studies that neglect the spatial dimension risk biased estimates of the effects of decentralization and intergovernmental transfers. This methodological insight underpins the empirical strategy adopted in the present paper.

## 3.4 Research Gaps

Despite the expanding literature, three gaps remain. First, no study has yet examined strategic spatial interactions among Cameroonian regional governments in the allocation of economic infrastructure expenditure. Previous work on Cameroonian decentralization has relied exclusively on non-spatial models. Second, while the original formulation of yardstick competition relies on residential mobility and competitive local elections, the top-down variant operating through central transfer policy has received limited empirical attention in Central Africa. Third, the intensity of spatial interactions appears to depend on a minimum level of own-source revenue, yet this dimension has not been explored in the Cameroonian context. The present paper addresses these gaps.

## 4. Research Gaps and Contribution

By providing the first spatial panel analysis of regional infrastructure spending in Cameroon, testing the top-down yardstick competition hypothesis in a Central African setting, and quantifying the differentiated effects of alternative financing instruments, this paper fills the gaps identified above and offers concrete policy implications for the design of intergovernmental transfers.

## 5. Data and Institutional Context

### 5.1 Nature and Source of Data

The data are drawn from the administrative accounts of the Cameroonian Public Treasury (CTP) and cover the ten regions of Cameroon at quarterly frequency from 2015Q1 to 2018Q4. The choice of quarterly frequency is motivated by the need to capture intra-annual budgetary dynamics and by data availability. Although the resulting balanced panel comprises only 160 observations (10 regions × 16 quarters), which is relatively modest, its structure and the use of the SPREGDPD estimator allow us to address the main econometric challenges endogeneity of the spatial lag, dynamics in infrastructure spending, and unobserved regional heterogeneity. The limited time span reflects the scarcity of consistent quarterly subnational fiscal data in Cameroon and, more broadly, in Central Africa.

The dependent variable, Infras_eco, measures regional capital expenditure on economic infrastructure (primarily roads: construction, rehabilitation and maintenance of regional and municipal roads). We use its natural logarithm to address skewness. The key explanatory variable of interest, Rp, captures own-source revenues collected by municipalities within each region (business tax, property taxes and local duties) and serves as a proxy for local fiscal autonomy. Control variables include four categories of intergovernmental transfers: the General Decentralization Grant (Dot_gen_dec), unconditional transfers from sectoral ministries (Trans_etat), city-to-district transfers (Dot_re_cu), and other subsidies (Autres_sub). All financial variables are expressed in natural logarithms.

## 6. Methodological Approach

### 6.1 The Spatial Autoregressive (SAR) Model

Following the literature on strategic interactions among local governments (Case et al., 1993; Revelli, 2006; Foucault et al., 2008), we estimate a spatial autoregressive (SAR) panel data model. In this specification, the dependent variable in region i depends both on its own characteristics and on the weighted average of the dependent variable in neighbouring regions.

where i indexes regions (i = 1, …, 10) and t indexes quarters (t = 1, …, 16); $Wy_t$ is the spatial lag; ρ is the spatial autoregressive coefficient (ρ > 0 indicates strategic complementarity); $\mu_i$ are regional fixed effects; $\lambda_t$ are time fixed effects; and $\varepsilon_{it}$ is the idiosyncratic error term.

### 6.2 Spatial Weighting Matrix

We employ a first-order contiguity matrix in which $w_{ij} = 1$ if regions i and j share a common border and 0 otherwise (diagonal elements are set to zero). The matrix is row-standardised so that each row sums to one, allowing ρ to be interpreted as the effect of the average neighbour. Contiguity is the most commonly used criterion in the literature on spatial interactions among territorial jurisdictions, particularly for road infrastructure whose spillovers propagate geographically.

### 6.3 Estimation Method: SPREGDPD (Arellano–Bond)

We estimate the model with the SPREGDPD estimator, which simultaneously addresses three econometric challenges: (i) the endogeneity of the spatial lag (the reflection problem of Manski, 1993), instrumented by lagged values of neighbours' expenditures; (ii) the dynamic nature of budgetary decisions, controlled by the inclusion of the lagged dependent variable; and (iii) regional fixed effects, eliminated by first-differencing (Arellano & Bond, 1991). Instrument validity is assessed with the Hansen over-identification test. Specification tests (Moran's I, LM-Lag versus LM-Error) support the choice of the SAR model over the SEM (see Appendix Tables A2 and A3).

## 7. Empirical Findings

### 7.1 Spatial Interaction in Economic Infrastructure

Table A4 reports the SPREGDPD estimates. The central result is a positive and statistically significant spatial autoregressive coefficient ρ, ranging from 0.0348 to 0.0436 across specifications. Economic infrastructure spending in a given region is therefore positively associated with that of its neighbours, consistent with the existence of horizontal strategic interactions among Cameroonian regions.

Although statistically significant, the magnitude of ρ remains modest by international standards (typically 0.10–0.40 in developed countries). This relatively weak interaction is consistent with the still limited fiscal autonomy of Cameroonian regions and their high dependence on central transfers (more than 80% of resources according to MINAT, 2023). It suggests that yardstick competition, while present, operates within a constrained institutional environment characterised by weak local tax bases and limited residential mobility. The estimated coefficient is nevertheless close to values reported for other African settings and aligns with the broader pattern that spatial interactions tend to be weaker in low- and middle-income countries than in high-income ones.

## 7.2 The Role of Own-Source Revenues

Own-source revenues (Rp) exert a positive and significant effect on economic infrastructure spending. In the full specification (Model 4), the coefficient is 0.0280 ($p < 0.05$), corresponding to a marginal effect of 2.8%. A 10% increase in own-source revenues is associated with a 0.28% increase in economic infrastructure expenditure. This result confirms the importance of local fiscal autonomy as a driver of productive investment and is consistent with earlier findings for Bolivia (Faguet, 2004) and Benin (Caldeira et al., 2012).

## 7.3 The Impact of the General Decentralization Grant

The General Decentralization Grant (Dot_gen_dec) exhibits a negative and highly significant coefficient (−0.1348, $p < 0.01$), implying a marginal effect of −13.5%. Regions receiving larger unconditional grants tend to spend less on economic infrastructure. This counter-intuitive result is consistent with the literature on vertical fiscal gaps (Rodden, 2004): heavy reliance on unconditional transfers weakens the link between local taxation and local spending, reduces accountability, and may encourage fiscal inertia or clientelist use of resources. Similar negative effects of unconditional transfers have been documented in other developing-country contexts. The finding constitutes one of the most policy-relevant results of the study and calls for a substantial redesign of Cameroon’s main equalisation instrument.

## 7.4 Analysis of Marginal Effects

Marginal effects reported in Table A5 allow a precise assessment of economic magnitudes. City-to-district transfers (Dot_re_cu) display the largest positive effect (27.4%), followed by other subsidies (15.6%) and own-source revenues (2.8%). The strong positive effect of city-to-district transfers suggests that vertically structured, intra-metropolitan transfers are better targeted toward productive investment than pure unconditional grants. By contrast, the large negative effect of the General Decentralization Grant (−13.5%) indicates that Cameroon’s primary equalisation instrument is currently associated with lower, rather than higher, economic infrastructure spending. This result reinforces the case for introducing performance-based components into the allocation formula and for strengthening local revenue mobilisation.

## 8. Policy Implications

Four operational recommendations follow from the empirical results.

### 8.1 Strengthening Local Fiscal Autonomy

The positive but modest effect of own-source revenues (marginal effect 2.8%) reflects the structural fragility of local fiscal autonomy in Cameroon, where more than 80% of municipal resources still depend on central transfers. Three concrete actions are required: (i) broadening the local tax base (business tax, property tax, residence tax) through the modernisation of taxpayer registers; (ii) improving collection efficiency via digitalisation and training of municipal tax agents; and (iii) reforming the allocation formula of the General Decentralization Grant so that it rewards municipalities that mobilise more own-source revenue relative to their potential.

### 8.2 Reforming the General Decentralization Grant

The negative effect of the General Decentralization Grant on economic infrastructure (−13.5%) is the most concerning finding of the study. A thorough overhaul is needed. We recommend that a significant share of the grant (for example 30%) be allocated on the basis of clear and verifiable performance criteria: own-source revenue mobilisation relative to potential, project completion rates, and the quality of realised infrastructure. Annual publication of allocation criteria and amounts received by each jurisdiction would enhance transparency and reduce the scope for clientelism.

### 8.3 Institutionalising Yardstick Competition through FEICOM

The positive spatial coefficient indicates that a “top-down” competitive mechanism already operates, albeit weakly and informally. The Special Fund for Equipment and Intercommunality (FEICOM) could institutionalise this mechanism by: (i) publishing an annual comparative dashboard of regional and municipal performance on economic infrastructure; (ii) conditioning part of its equipment grants on year-on-year improvement in selected indicators; and (iii) organising regular knowledge-sharing meetings so that competition is complemented by collective learning.

### 8.4 Targeting the Poorest Regions

The poorest regions (Adamawa, East, Far North, North) are likely to benefit least from yardstick competition because they lack a critical mass of own-source revenue. Specific instruments are therefore required: multi-year seed grants for jurisdictions below a minimum own-source revenue threshold, reinforced technical assistance, and, in the longer run, a horizontal equalisation mechanism financed by wealthier jurisdictions.

## 9. Conclusion

This study has examined spatial interactions among Cameroon's ten regions in the allocation of economic infrastructure expenditure in order to test the hypothesis of "top-down" yardstick competition. Using the SPREGDPD estimator and quarterly panel data for 2015–2018, we obtain three main results: (i) a positive and significant spatial autoregressive coefficient ($\rho \approx 0.035$–$0.044$), consistent with the existence of strategic horizontal interactions; (ii) a positive effect of own-source revenues (marginal effect 2.8%), confirming the role of local fiscal autonomy; and (iii) a negative effect of the General Decentralization Grant (marginal effect −13.5%), suggesting crowding-out or inefficient allocation of unconditional transfers.

By providing the first spatial econometric evidence on strategic interactions among Cameroonian regional governments, extending the analysis of top-down yardstick competition to Central Africa, and quantifying the differentiated impact of alternative financing instruments, the paper contributes to both the academic literature and the policy debate on fiscal decentralization. The results point to four priority reforms: strengthening local revenue mobilisation, redesigning the General Decentralization Grant, institutionalising performance comparison through FEICOM, and introducing targeted support for the poorest regions.

Several extensions remain for future research. Testing for spatial heterogeneity in the intensity of yardstick competition (according to regional wealth or proximity to the capital), extending the panel beyond 2018 to assess the impact of the 2019 General Code, and conducting comparative analyses with other CEMAC countries would further enrich the understanding of spatial fiscal interactions in Central Africa.

## ppendix A, Tables

### Table A1, Variable Descriptions

| Variable | Description | Source |
|---|---|---|
| Infras_eco | Total investment in economic infrastructure (mainly roads) | CTP |
| Dot_gen_dec | General Decentralization Grant (state transfer to LGUs) | CTP |
| Trans_etat | Unconditional state transfers to LGUs | CTP |
| Rp | Own-source revenues from local taxation | CTP |
| Dot_re_cu | Financial transfers from city hall to district municipalities | CTP |
| Autres_sub | Other government subsidies (earmarked taxes via FEICOM or Treasury) | CTP |

*Source: Cameroonian Public Treasury accounts (CTP), 2015–2018,*

**Table A2, Specification Tests-Moran's I**

| Test | Statistic | p-value | Decision |
|---|---|---|---|
| Moran's I (OLS residuals) | -0,0727 | 0,087 | Reject $H_0$ at 10% |
| Moran's I (full SAR model) | 0,0004 | 0,05 | Fail to reject $H_0$ |

*Note: Moran's I on OLS residuals (with fixed effects, no spatial term) is negative and significant at 10% (p = 0,087), indicating that a spatial specification is necessary, After estimating the SAR model, Moran's I on residuals is no longer significant (p > 0,05), confirming that the spatial autoregressive term correctly captures spatial dependence,*

*Source: Author,*

**Table A3, LM Tests for Choosing Between SAR and SEM Models**

| Test | Statistic | p-value |
|---|---|---|
| LM-Lag (SAR) | 100,75 | 0,000 |
| LM-Error (SEM) | 23,36 | 0,000 |
| RLM-Lag (Robust) | 30,50 | 0,000 |
| RLM-Error (Robust) | 20,48 | 0,000 |

*Note: Both LM-Lag and LM-Error tests are significant, The higher LM-Lag statistic (100,75 vs, 23,36) and the significant robust RLM-Lag confirm the preference for the SAR specification,*

*Source: Author,*

## Table A4, Econometric Estimation Results

| Dependent variable: infras_eco (log) | | | | |
|---|---|---|---|---|
| **SPREGDPD = Spatial Panel Linear Dynamic Regression (Arellano–Bond)** | | | | |
| | **(1)** | **(2)** | **(3)** | **(4)** |
| W1y_infras_eco | 0,0436*** | 0,0349** | 0,0372** | 0,03481** |
| | (0,0219) | (0,0165) | (0,0161) | (0,0176) |
| Rp | 0,0220 | 0,0236* | 0,0227* | 0,0280** |
| | (0,0167) | (0,0128) | (0,0125) | (0,0137) |
| Autres_sub | — | 0,1978*** | 0,1871** | 0,1557*** |
| | | (0,0123) | (0,0125) | (0,0140) |
| Dot_re_cu | — | — | 0,1303*** | 0,2737*** |
| | | | (0,0428) | (0,0487) |
| Dot_gen_dec | — | — | — | -0,1348*** |
| | | | | (0,0125) |
| Constant | 5,137** | 5,982*** | 3,729** | 3,936** |
| | (1,675) | (1,235) | (1,440) | (1,569) |
| Observations | 160 | 160 | 160 | 160 |
| **Test of overall model validity** | | | | |
| Wald test | 520,43*** | 1046,78*** | 119,73*** | 1106,43*** |
| F-test | 173,48*** | 261,69*** | 238,75*** | 184,40*** |
| $R^2$ | 0,9998 | 0,8776 | 0,8917 | 0,8848 |
| Adjusted $R^2$ | 0,9998 | 0,9657 | 0,8887 | 0,8808 |
| **Spatial autocorrelation tests** | | | | |
| Global Moran MI | -0,0727* | -0,1052** | -0,1054** | 0,0004*** |
| Global Geary GC | 0,6310 | 0,9380** | 0,9388** | 0,8700*** |
| Global Getis-Ord GO | 0,2470 | 0,3578*** | 0,3585*** | -0,0013*** |
| **Heteroskedasticity tests** | | | | |
| LM test | 100,75*** | 23,36*** | 30,50*** | 20,48*** |
| White test (BPG) | 100,75*** | 141,16*** | 166,68*** | 170,79*** |
| Wald test (het,) | 47,66*** | 57,63*** | 75,25*** | 50,52*** |
| **Residual normality tests** | | | | |
| Jarque–Bera LM | 3299,85*** | 206,44*** | 234,07*** | 59,46*** |
| White IM test | 3311,88*** | 214,34*** | 240,31*** | 70,19*** |
| Small LM Skewness | 65,31*** | 14,42*** | 20,49*** | 6,21** |
| Small LM Kurtosis | 57,20*** | 26,16*** | 27,07*** | 15,26*** |

*Notes: Standard errors in parentheses, W1y_infras_eco = spatial weight matrix W × dependent variable y, *** p < 0,01, ** p < 0,05, * p < 0,1, Source: Author,*

**Table A5, Marginal Effects Analysis (Model 4)**

| Variable | Marginal Effect |
|---|---|
| ρ (Spatial interaction) | 0,0348 (3,48%) |
| Rp (Own-source revenues) | 0,0280 (2,80%) |
| Autres_sub (Other grants) | 0,1557 (15,57%) |
| Dot_re_cu (City-to-district transfers) | 0,2737 (27,37%) |
| Dot_gen_dec (General Decentralization Grant) | -0,1348 (-13,48%) |

*Note: Marginal effects interpreted as elasticities — percentage change in economic infrastructure associated with a 1% change in each explanatory variable, all else equal,*

*Source: Author,*